\documentclass[prl,aps,showpacs,groupedaddress,twocolumn,nofootinbib]{revtex4}

\usepackage{amsmath,amsfonts,amssymb}
\usepackage{wrapfig}
\usepackage{graphicx}
\usepackage{bbm}
\usepackage{graphics}

\def \beq {\begin{equation}}
\def \eeq {\end{equation}}
\def \ba {\begin{eqnarray}}
\def \ea {\end{eqnarray}}
\def \a {\hat{a}}
\def \ad {\hat{a}^{\dag}}

\newcommand{\ketbrad}[1]{|#1\rangle\!\langle #1|}
\newcommand{\mb}[1]{\mbox{\boldmath$#1$}}

\def\ket#1{\left| #1\right>}
\def\bra#1{\left< #1\right|}

\renewcommand{\section}[1]{\paragraph{#1---}}
\renewcommand{\subsection}[1]{}
\begin{document}

\title{Wigner crystals of ions as quantum hard drives}
\author{J. M. Taylor$^1$ and T. Calarco$^2$}
\affiliation{$^1$Department of Physics, Massachusetts Institute of Technology, Cambridge, MA  02139\\
$^2$Department of Physics, Harvard University, and ITAMP, Cambridge, MA  02138}
\date{\today}
\begin{abstract}
  Atomic systems in regular lattices are intriguing systems for
  implementing ideas in quantum simulation and information processing.
  Focusing on laser cooled ions forming Wigner crystals in Penning
  traps, we find a robust and simple approach to engineering
  non-trivial 2-body interactions sufficient for universal quantum
  computation.  We then consider extensions of our approach to the
  fast generation of large cluster states, and a non-local
  architecture using an asymmetric entanglement generation procedure
  between a Penning trap system and well-established linear Paul trap
  designs.
\end{abstract}
\maketitle

Quantum information processing using trapped ions has been the focus
of theoretical~\cite{ciracZollerPhysToday} and
experimental~\cite{schmidtkaler03,blinov04,wineland05,langer05,boyd06}
efforts over the past decade.  The coherence times of ions can exceed
seconds, while manipulation and entanglement time scales can be
as fast as tens of microseconds.  So far, approaches to
scaling these systems to many ions have met with significant issues,
both in linear Paul trap systems, where increasing numbers of ions
leads to control difficulties, and in proposed more complex trap arrays, where
``shuttling'' of quantum information using gate electrodes would allow
for a scalable architecture~\cite{kielpinski02}. A possible solution
is to separate the processing elements (processor qubits) from the
memory~\cite{oskin02}.

One natural system to consider as a quantum memory is a Wigner crystal
of ions in a Penning trap~\cite{RevModPhys.71.87}.  Such crystals can
be robustly formed~\cite{jensen:033401}, and are dynamically stable,
with tens of thousands of ions in a given trap.  In addition, the
strength of the Coulomb interaction leads to large separations between
individual ions, making individual addressing of ions in such lattices
a distinct possibility, in contrast to present control in neutral atom
and polar molecule lattices~\cite{bloch05,buchler07}.

In this Letter we develop an approach to quantum memory and
entanglement generation that takes full advantage of the advances in
ion trap technology for building large Wigner crystals of ions in
Penning traps.  Using a modulated-carrier ``push'' gate adapted from
linear ion trap quantum computing
schemes~\cite{sorensen99,calarco01,garcia-ripoll03,zhu06}, we find a
fast but adiabatic method for building small clusters of entanglement
which is insensitive to thermal phonons in 2D and 3D Wigner
crystals. We take advantage of some of the unique features of Penning
traps, such as rotation of the crystal, to provide simplifications in
the necessary hardware to implement these ideas in 2D Wigner crystals.
We further show that such a quantum memory device can also be used
directly for cluster state quantum computation.
Our approach follows recent work
\cite{porras06} on performing quantum gates in 2D Wigner
crystals.  Finally, non-deterministic entanglement generation between
distant ions suggests a processor (linear Paul trap) and memory (2D
Wigner crystal) architecture based upon a quantum register
approach~\cite{oskin02,jiang07}, where the low photon collection
efficiency from ions in the memory is offset by an asymmetric
entanglement generation scheme using a weak cavity coupled to ions in
the processor~\cite{mundt02,blinov04}.

\begin{figure}
\centering
\includegraphics[width=1.8in]{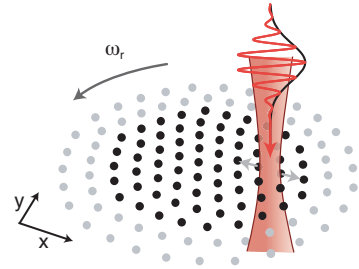}
\caption{Two-qubit gate via intensity modulation of a laser
addressing a pair of ions in a crystal rotating at frequency
$\omega_r$.\label{f:crystal}}
\end{figure}

We start by considering a Wigner crystal of ions, rotating
in a Penning trap~\cite{RevModPhys.71.87} with harmonic confinement
with frequencies $\omega_{xy}$ (in the lateral directions) and
$\omega_z$ (in the vertical direction).  With characteristic ion
spacings $d \sim 10\ \mu$m, tightly focused lasers allow for
individual addressing of ions (see Fig.~\ref{f:crystal}).
Laser cooling can reduce the temperature $\approx 1$ mK, yielding on
the order of $10^2 - 10^3$ phonons in the softest (lateral) modes.  By
using long-lived, metastable states of the ions as a quantum memory,
we may neglect memory errors in our discussion.  A tightly-focused
laser allows for nearest-neighbor phase gates and for single ion
operations.  Large-scale computation may be considered using either
nearest-neighbor couplings or via a variety of quantum communication
techniques developed for quantum repeater protocols.  When used in
conjunction with the deterministic phase gate developed below and
local single ion operations (performed, e.g., via Raman transitions),
this will suffice for performing large scale quantum
algorithms~\cite{jiang07} by using the remote CNOT
gate~\cite{gottesman99}.

\section{Modulated-carrier gate}

A spatially inhomogeneous laser detuned from the appropriate
transitions between internal (qubit) states of an ion (a two-level
system with Pauli matrices $\sigma_i^{x,y,z}$) produces a
ponderomotive force $\vec{f}_i$ due to the gradient in its intensity.
Using an appropriate combination of polarizations and frequencies, in
analogy with alkali atoms \cite{Jaksch}, the sign of the force becomes
dependent upon the internal state of the ion, with the associated
perturbation to the system:
\begin{equation}
 V  = \sum_i \left[\vec{x}_i \cdot \vec{f}_i(t)\right] \sigma^z_i \label{e:force}
\end{equation}
where $\vec{x}_i$ is the displacement of ion $i$ away from its
equilibrium position. The latter can take place either along the
separation between two individual ion microtraps \cite{calarco01} or
perpendicularly to the plane of an ion Wigner crystal in a Penning
trap \cite{porras06}. In both schemes, as the ion displacements are
coupled (via phonons), such a push leads to an effective $\sigma^z_i
\sigma^z_j$ interaction. Adiabaticity is required for vibrational
excitations to be absent after the gate. This bounds the clock speed
to be lower than the frequency of trapping in the push direction:
tight traps are needed for fast, temperature-insensitive operation.

\subsection{Carrier push derivation}

We now introduce a simple variant of the fast-kick ``push'' gate which
allows us to use even the soft (lateral) modes when their temperature
is extremely high.  Our variant uses slow modulation of a fast,
oscillating state-dependent force.
The oscillation averages any ion motion to zero over the course of the
gate, while the in-phase oscillation of nearby ions leads to a
non-trivial phase evolution and the desired quantum gate between ions
in the crystal.  In addition, as our gate allows for non-trivial
oscillation of ion positions in all three spatial dimensions (versus
only in the tightly confined direction for the vertical gate), it can
work using a single laser beam and in three dimensional crystals.

It is instructive to recall the general description of ``push'' phase
gates when in a complex crystal~\cite{porras06,garcia-ripoll03}. We
start by rewriting the Hamiltonian of $N$ interacting ions to second
order in displacement from the equilibrium positions $H = \sum_K \hbar
\omega_{K} \ad_{K} \a_{K}$, using normal-mode coordinates indexed by
$K = \{ \vec{k}, \lambda\}$ (the wavevector and polarization),
$\vec{x}_i = \sum_K M_{iK} \vec{e}_{K}(\alpha_K/\sqrt{2}) (\a_K +
\ad_K)$. 
The $\alpha_K = \sqrt{\hbar/m \omega_K}$ are the oscillator ground
state lengths; the matrix $M$ is orthogonal ($M^t M = M M^t = 1$).
The perturbation $V$
can now be written as
\begin{equation}
    V = \sum_K \alpha_K f_K(t) (\ad_K + \a_K)/\sqrt{2},
\end{equation}
where $f_K(t)$ is the state-dependent force on normal mode $K$
defined via the transformaton $M$ and Eqn.~\ref{e:force}.

The problem factorizes into $3 N$ independent, driven oscillators.
For scenarios with ${\rm lim}_{t \rightarrow \pm \infty} f(t) = 0$, the oscillator evolution is given by the unitary transform $U_K(t)
= e^{-i \phi_K(t)} \exp(\beta_K \ad_K - \beta_K^* \a)$, where
$\phi_K$ and $\beta_K$ satisfy the differential
equations~\cite{garcia-ripoll03}
\begin{equation}
\dot{\beta}_K = -i \omega_K \beta_K + i \frac{\alpha_K}{\hbar
\sqrt{2}} f_K(t), \; \dot{\phi}_K = \frac{\alpha_K}{\hbar \sqrt{2}}
f_K(t) {\mathcal Re}[\beta_k(t)]\
\end{equation}
which are exact to second order.

We now seek an approach which still maintains no net change in
displacement and no dependence of the overall phase on phonon state,
but can operate on time scales on the order of $\omega_K$.  We add a
sinusoidal variation to the force [$f(t) \rightarrow \cos(\nu t)
f(t)$].  The carrier frequencies $\nu$ must be fast with respect to
$\omega_K$; qualitatively, this averages out any net displacement.
If the modulation $f(t)$ is slow as compared to $\nu$ (but with no
restriction with respect to $\omega_K$), we can perform a similar
adiabatic elimination as above, and get a gate with the same
desirable properties that can operate non-trivially on arbitrarily
``soft'' phonon modes at very high tempeatures.

For adiabatic elimination with respect to $\nu$, we choose the ansatz
$\beta = \beta_+ e^{i \nu t} + \beta_- e^{-i \nu t}$ (subscripts
omitted for clarity).  Setting $\dot{\beta} = 0$
yields $\beta_{\pm} = \alpha f(t)/[2 \sqrt{2} \hbar(\omega \pm
\nu)]$.  We find the displacement of a normal mode induced by the
gate is proportional to the force applied, and can be made zero
independent of initial phonon state by starting and ending with zero
force. This eliminates any potential error due to entanglement
between phonons and the internal states of the ions.

We now examine the two-body phase induced in this new
scenario.  The differential equation for phase is now:
\begin{equation}
  \dot{\phi} = \frac{\alpha^2}{2 \hbar^2} f^2(t) \frac{\omega}{(\omega^2-\nu^2)}
  \cos^2(\nu t).
\end{equation}
Averaging the quickly varying component lets us replace $\cos^2(\nu
t)$ with $1/2$. Returning the mode index, $K$, we find the overall
phase accumulated, $\sum_K \phi_K(T)$, for a gate occurring over a
time 0 to $T$ does not depend on the phonon initial state. However,
the internal states of the ions are affected by the unitary $\exp(-i
\sum_{ij} \phi_{ij} \sigma^z_i \sigma^z_j)$ where the two-body
phases are given by
\begin{equation}
  \phi_{ij} = \sum_{\lambda} S^\lambda_{ij}
  \int_0^T (\vec{f}_i(t) \cdot \vec{e}_{\lambda}) (\vec{f}_j(t) \cdot
  \vec{e}_{\lambda}) dt. \label{e:adiabatic}
\end{equation}
The pulse-shape
independent form factor is
\begin{equation}
 S^\lambda_{ij} = -\sum_k \frac{\alpha_{k,\lambda}^2
   \omega_{k,\lambda}}{4\hbar^2(\nu^2-\omega_{k,\lambda}^2)} M_{ik,\lambda}M_{jk,\lambda}\ \label{e:Sfastall}
\end{equation}
(the polarization vectors $\vec{e}_K$
only depend on $\lambda$).

Expanding in inverse powers of the large carrier frequency $\nu$, we
note that the first term, $O(\nu^{-2})$, is proportional to $\sum_k
M_{ik,\lambda} M_{jk,\lambda} = 0$ (due to the orthogonality of the
matrix $M$).
The first non-zero term is $O(\nu^{-4})$.
Compared to adiabatic gates, this modulated-carrier push gate is
inverted in sign and reduced in phase by a factor
$(\omega/\nu)^4/2$. For lateral gates, $\omega \sim \omega_{xy}$ is a
characteristic confinement energy for a single ion in the crystal, and for vertical gates, $\omega \sim \omega_z$.

\section{Performance}

\begin{figure}
\centering
\includegraphics[width=3.3in]{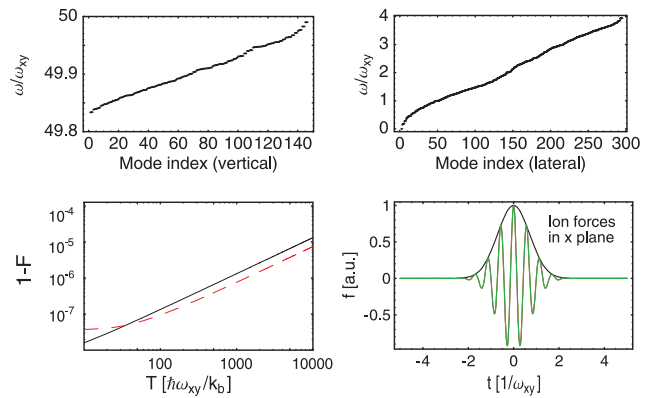}
\caption{ (top) Phonon spectrum for vertical and lateral phonons for
a trap with $N=147$ ions and $\omega_z / \omega_{xy} = 50$. (bottom)
Fidelity versus temperature for the vertical-phonon-mode
gate of Ref.~\cite{porras06} (red-dashed line) and for the fast
carrier gate with $\nu = 11 \omega_{xy}$ between the center-most
pair of ions.  Left: fast carrier gate's and vertical gate's forces
on one of the two ions over the gate time; both gates operate in a
time $\tau \sim 1/\omega_{xy}$. Anharmonic corrections to the
fidelity are not included here. For this choice of parameters, the
vertical gate \cite{porras06} requires 20 times the force (and laser power) of our
fast carrier gate to achieve the same final $\pi$ two-body phase.
\label{f:gates}}
\end{figure}

We performed numerical simulations of the modulated-carrier gate's
performance for finite-size 2D and 3D Wigner crystals ($N=147$ shown
in Fig.~\ref{f:gates}) to compare to the equivalent
adiabatic gate and the proposed vertical gate of
Ref.~\onlinecite{porras06}. Our simulations minimized the classical energy
of the ions in a Penning trap to determine the equilibrium
configuration.  Then, expanding to second order in displacements from
equilibrium, the normal mode coordinates were found.  Coriolis forces
were neglected; their inclusion does not qualitatively change our
results.  A gate between the two centermost ions was simulated by
computing the displacement $\beta_K$ and phase $\phi_K$
for each phonon mode. The fidelity was calculated by considering the
overlap of the final state with the desired state, traced over the
phonon degrees of freedom and minimized over all possible initial
states.  

We find that for the same physical parameters, the ratio of
forces (i.e., laser power) required for achieving a $\pi$ phase for
the vertical gate and the fast carrier gate goes as $(\omega_z /
\nu)^2$, consistent with the ratio between adiabatic and fast carrier
gates derived above.  Thus, the fast carrier gate requires
substantially less laser power for the same conditions with negligible
reduction in fidelity.  Alternatively, the gate time could be reduced,
enhancing the overall performance of quantum information
protocols. For specificity, setting $\omega_{xy} = 200$ kHz, $\omega_z
= 10$ MHz, and a gate time $\tau = 5\ \mu$s, we find $\nu=2.2$ MHz
provides $1 - F< 10^{-5}$ with negligible heating. Even smaller errors are found in
simulations of the 3D crystal under the same approximations.  

A practical limitation occurs due to the spontaneous emission induced by the off-resonant laser interactions.  Tight focusing increases the force for the same laser power; thus, using a pair of adjacent, narrow-waist ($\lesssim 2\ \mu$m) laser beams reduces spontaneous emission and power requirements.  For specificity, using a transition with spontaneous emission of $\gamma = 20$ MHz and lasers with peak Rabi frequency of 100 GHz detuned 100 THz from the atomic transition, a laser power of $\sim 3$ mW per beam is required for our gate, with an induced error of $\lesssim 0.1\%$ per gate.

\section{Quantum cluster state generation}

We now consider an approach that takes advantage of the Coulomb
interactions in the lattice to create and dynamically extend a cluster state
for universal measurement-based computation.  Specifically, the goal is to obtain a weighted-graph state
$\exp({\rm i}\sum_{ij}\sigma_i^z \sigma_j^z \vartheta_{ij}/2 )
\ket{+,\dots,+}, $
where in the ideal case $\vartheta_{ij}$ equals $\pi$ between nearest neighbors
on a square lattice, and zero otherwise. On a triangular lattice
like the one available in many-ion Penning traps, this can be achieved
if $\vartheta_{ij}$ is made to vanish along one side of each lattice
cell, and to be $\pi$ on the other two. The idea is to obtain this
via a global $\pi/2$ qubit rotation followed by a push gate acting on all three cell vertices at the same time,
possibly with a laser swept at constant velocity through the cell
itself, to take advantage of the uniform circular motion of the
lattice.
\begin{figure}
\centering
\includegraphics[width=1.5in]{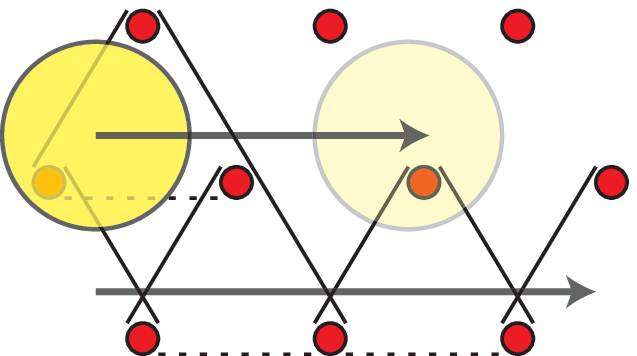} 
\includegraphics[width=1.5in]{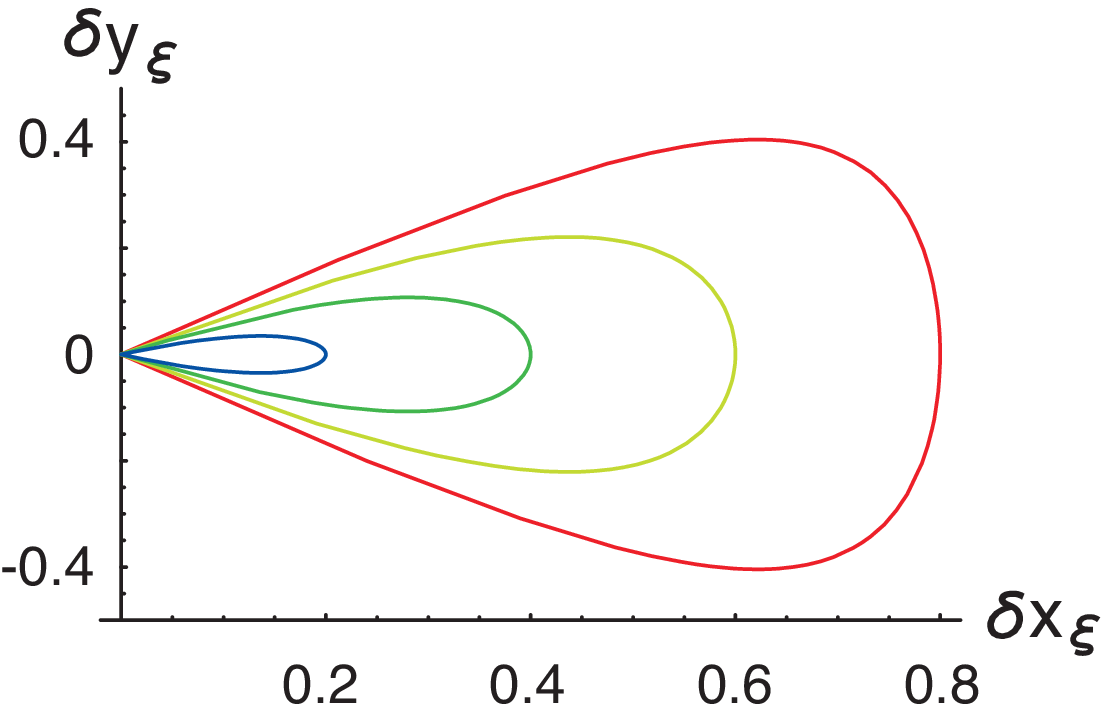}
\caption{ \label{f:sweep} (color online) a) A laser is swept
adiabatically from left to right, leading to a weighted graph state
with different phases for solid and dashed lines. b) Laser
displacements in the crystal plane needed to obtain a constant sweep
velocity at given distances $\xi/R=0.2,0.4,0.6,0.8$ (from blue to
red) from the center, in the rotating crystal's frame.
}
\end{figure}

We start by considering a focused laser beam of waist $\sigma$ (in
units of the lattice length $d$) adiabatically swept at constant
velocity $v$ through the Wigner crystal, along a direction parallel
to one of the lattice vectors (Fig.~\ref{f:sweep}), at half the
height of a triangular cell. The effect of this sweep is, apart from
a global single-qubit rotation, to generate a weighted-graph phase, where $\vartheta_{ij}$ takes value
$\varepsilon\theta(\omega)$ on the cell side that is parallel to the
sweep direction, and $\theta(\omega)$ on the other two sides, with
$\varepsilon=e^{-3/(8\sigma^2)}(11-8\sigma^2)/(\sigma^2+8)$, while
\begin{equation}
\theta(\omega)=\frac{\Omega_0^4}{\omega^2\Delta^2}\frac{\alpha^4}{d^4}
\frac{q^2}{\hbar\epsilon_0v}\frac{e^{-1/(2\sigma^2)}}{\sqrt{8\pi}\sigma}
\left(\frac 1{\sigma^2}+\frac 18\right),\label{theta}
\end{equation}
where $\alpha=\sqrt{\hbar/(m\omega)}$, $\Omega_0$ is the peak Rabi
frequency corresponding to the center of the laser beam, $\Delta$ is
its detuning from the ion's internal transition, and $q$ is the
electron charge. 
Using the
fast carrier modulation described above, the
semiclassical calculation of Eq.~(\ref{theta}) is no longer valid, but
the discussion of Eq.~(\ref{e:Sfastall}) shows that the resulting
phase is simply $-\theta(\nu)/2$. A cluster state is then obtained by
making $\varepsilon$ small via an appropriate choice of the laser waist (numerically,
$\sigma \lesssim 0.2$), while tuning $\theta(\nu)/2$ to $\pi$ by
adjusting the other experimental parameters such as laser power. 

Care
needs to be taken to ensure a sweep having a given distance $\xi$ from
the trap center and velocity $v$ in the rotating crystal's frame. To
this end we apply to the laser, initially focused at a distance $R$
from the center, a displacement $\delta \mb d_\xi(t)=\{\delta
x_\xi(t),\delta y_\xi(t)\}$ of the form
\begin{eqnarray}
\delta x_\xi(t)&=&[R-\xi/\cos(\omega_r t)]\Theta(\chi-|\omega_r t|)\;,\\
\delta y_\xi(t)&=&\xi[\omega_r t \tan(\chi)/\chi-\tan(\omega_r
t)]\Theta(\chi-|\omega_r t|)\;\quad
\end{eqnarray}
(see Fig.~\ref{f:sweep}), where $\chi\equiv\arccos(\xi/R)$.
Cluster state generation as presented uses a single laser beam, resulting in substantially higher laser power requirements than the two qubit gate with two beams as described above.  In particular, for errors per gate $\lesssim 0.1\%$, a detuning of 200 THz and peak Rabi frequency of 4 THz (corresponding to 5 W of laser power) would be required.  More complex laser motion could reduce these requirements.

\section{Asymmetric entanglement generation}

\begin{figure}
\centering
\includegraphics[width=3.0in]{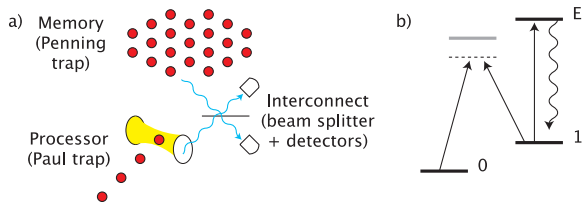}
\caption{ \label{f:cycling} a) Schematic of a quantum processor
(such as a linear Paul trap) coupled via a high finesse cavity to a
photodetector system to allow interference with photons from an ion
Wigner crystal in a Penning trap nearby. b) Ion level structure
}
\end{figure}

We conclude with a brief discussion on the implementation of
circuit-based computation using entanglement generation and remote
CNOTs.  We use a quantum processor unit (such as a linear Paul trap)
separated from the quantum memory unit (our Wigner-crystal-based
quantum hard drive -- see Fig.~\ref{f:cycling}), characterized by
photon collection efficiencies $\eta$ and $\eta'$ respectively.
Without loss of generality, we will assume $\eta'>\eta$, as it can be
achieved via coupling with high finesse
cavities~\cite{mundt02,blinov04}.
Our two-click asymmetric entanglement generation procedure starts with
an equally weighted superposition $\ket{+} = \ket{0} + \ket{1}$.  An
optical $\pi$ pulse produces photons via spontaneous emission at a rate $\gamma$ from the
$\ket{1}\leftrightarrow\ket{E}$ transition.  Then the photons are interfered on a beam
splitter. Without assuming photon number resolving detectors, the
state after one ``click'' becomes
\beq \eta' \ket{\psi_\pm}\bra{\psi_\pm} + O(\sqrt{\eta'})
\ketbrad{11}, \eeq with
$\ket{\psi_\pm}\equiv\sqrt{\eta/\eta'}(\ket{01}\pm\ket{10})$.  To
symmetrize the entangled state and simultaneously remove the
$\ket{11}$ component, a $\pi$ pulse between the metastable states
($\ket{1} \leftrightarrow \ket{0}$) followed by repetition of the
above protocol results in a pure state $\ket{\Phi^+} =
(\ket{01}\pm\ket{10})/\sqrt{2}$.  The overall procedure succeeds with
probability $\eta \eta'$, indicating that the time required is
$(\gamma \eta \eta')^{-1}$. A standard one-click scheme
\cite{childress05} with excitation probability $p$ takes a time
$(\gamma \eta p)^{-1}$, and succeeds with error rate $O(p)$, i.e., the
higher fidelity a pair one wishes to generate, the longer it takes. By
contrast, in our scheme the fidelity can be high without a further
increase in generation time.

Thus, for large-scale computation, a central processor unit with high
collection efficiency allows for high-fidelity gates between elements
of the ``hard drive'' memory on a timescale $2/\Gamma \eta \eta'$ (see
Ref.~\onlinecite{jiang07} for further improvements). For concreteness,
we take a radiative decay rate of $\Gamma = (2 \pi) 10$ MHz, $\eta =
10^{-3}$ (confocal approach with low numerical aperture lens), and
desired infidelity $1-F < 10^{-4}$. Entanglement generation between
two such ions would take a time $\sim 10$ ms or longer; in contrast,
for $\eta' \sim 0.1$, using the intermediate quantum processor leads
to entanglement generation between processor and both ions in a time
of order 100 $\mu$s, comparable to the phase gate operation times
already discussed.

This complements the quantum hard drive architecture described above,
providing a comprehensive toolbox for universal quantum computation
with ion crystals in Penning traps that relies on existing
technologies under available experimental conditions.

\begin{acknowledgments}

We thank J. Bollinger and D. Porras for helpful discussions.  JMT is
supported by Pappalardo, and TC by the European
Commission through projects SCALA and QOQIP.

\end{acknowledgments}

\bibliographystyle{apsrev}
%

\end{document}